\documentclass[twocolumn,showpacs,preprintnumbers,amsmath,amssymb]{revtex4}

\usepackage{bm}% bold math
\usepackage{texdraw}

\def\dv{\mathop{\rm div}\nolimits}
\def\curl{\mathop{\rm curl}\nolimits}
\def\skw{\mathop{\rm skw}\nolimits}
\def\sym{\mathop{\rm sym}\nolimits}
\def\tr{\mathop{\rm tr}\nolimits}

\newcommand{\GG}{\ensuremath{\mathbf G}}
\newcommand{\II}{\ensuremath{\mathbf I}}
\newcommand{\LL}{\ensuremath{\mathbf L}}
\newcommand{\MM}{\ensuremath{\mathbf M}}
\newcommand{\QQ}{\ensuremath{\mathbf Q}}
\newcommand{\Sb}{\ensuremath{\mathbf S}}
\newcommand{\WW}{\ensuremath{\mathbf W}}
\newcommand{\ab}{\ensuremath{\mathbf a}}
\newcommand{\ee}{\ensuremath{\mathbf e}}

\newcommand{\mm}{\ensuremath{\mathbf m}}
\newcommand{\nn}{\ensuremath{\mathbf n}}
\newcommand{\rr}{\ensuremath{\mathbf r}}
\newcommand{\uu}{\ensuremath{\mathbf u}}
\newcommand{\vv}{\ensuremath{\mathbf v}}
\newcommand{\bz}{\ensuremath{\mathbf 0}}
\newcommand{\epp}{\ee_+}
\newcommand{\emm}{\ee_-}
\font\filt=msbm10 
\newcommand{\filR}{\hbox{\filt R}}
\newcommand{\filS}{\hbox{\filt S}}
\newcommand{\bnu}{\mbox{\boldmath $\nu$}}
\newcommand{\btau}{\mbox{\boldmath $\tau$}}
\newcommand{\kuno}{{\kappa_1}}
\newcommand{\kdue}{{\kappa_2}}
\newcommand{\ed}{{{\bf e}_2}}
\newcommand{\et}{{{\bf e}_3}}
\newcommand{\lad}{{\lambda_2}}
\newcommand{\lat}{{\lambda_3}}
\newcommand{\tp}{\otimes}
\newcommand{\vp}{\wedge}
\newcommand{\grad}{{\nabla}}

\begin{document}

\title{Bulk and surface biaxiality in nematic liquid crystals}

\author{Paolo Biscari, Gaetano Napoli, Stefano Turzi}
\affiliation{Dipartimento di Matematica. Politecnico di Milano\\
Piazza Leonardo da Vinci, 32. 20133 Milan, Italy}
\email{paolo.biscari@polimi.it}

\date{\today}

\begin{abstract}
Nematic liquid crystals possess three different phases: isotropic,
uniaxial, and biaxial. The ground state of most nematics is either
isotropic or uniaxial, depending on the external temperature.
Nevertheless, biaxial domains have been frequently identified,
especially close to defects or external surfaces. In this paper we
show that any spatially-varying director pattern may be a source of
biaxiality. We prove that biaxiality arises naturally whenever the
symmetric tensor $\Sb=(\grad \nn)(\grad \nn)^T$ possesses two
distinct nonzero eigenvalues. The eigenvalue difference may be used
as a measure of the expected biaxiality. Furthermore, the
corresponding eigenvectors indicate the directions in which the
order tensor \QQ\ is induced to break the uniaxial symmetry about
the director \nn. We apply our general considerations to some
examples. In particular we show that, when we enforce homeotropic
anchoring on a curved surface, the order tensor become biaxial along
the principal directions of the surface. The effect is triggered by
the difference in surface principal curvatures.

\end{abstract}


\maketitle

Nematic liquid crystals are aggregates of rod-like molecules. Early
theories \cite{33os,33zoch,58fr} used a single order parameter, the
\emph{director\/}, a unit vector pointing along the average
microscopic molecular orientation. Most nematic phenomena fit well
within the classical description. However, the transition from
ordered to disordered states escapes the director theory. The
classical microscopic description of defects and surface phenomena
yields undesired results as well. The order-tensor theory put
forward by de Gennes \cite{69dg,95dgpr} focuses on the orientational
probability distribution, and introduces the measures of the degree
of orientation and biaxiality. Within this theory, a nematic liquid
crystal possesses three different phases, which can be identified
through their optical properties, since its Fresnel ellipsoid is
closely related to the order tensor itself \cite{94virga}. A
\emph{isotropic\/} liquid crystal is characterized by an isotropic
order tensor, and optically behaves as an ordinary fluid. A
\emph{uniaxial\/} nematic possesses a unique optic axis. Its order
tensor has two coincident eigenvalues. Finally, in a
\emph{biaxial\/} nematic the eigenvalues of the order tensor are all
different, and the Fresnel ellipsoid possesses two optic axes.

Within the Landau-de Gennes theory, the ground state may be either
isotropic or uniaxial, depending on the external temperature.
However, biaxial domains have been predicted and observed,
especially close to defects and external boundaries. Schopohl and
Sluckin \cite{87shsl} analyzed in detail the biaxial core of a
$+\frac{1}{2}$ nematic disclination. More recent studied show that a
biaxial \emph{cloud\/} surrounds most nematic defects \cite{97bipe},
and both analytic \cite{01robr,03bisl} and numeric
\cite{02chife,05chipa} asymptotical descriptions of biaxial defect
cores have been derived. Other examples of defect-induced biaxiality
involve integer-charged disclinations \cite{97bivi,99krvizu,02krvi}
and cylindrical inclusions \cite{05mcvi}. The onset of surface
biaxiality is closely related to the presence of a symmetry-breaking
special direction, which coincides with the surface normal
\cite{94bicavi}. Indeed, biaxiality has been predicted close to both
external boundaries \cite{95mate,99grdi} and internal
isotropic-nematic interfaces \cite{93chen,97pnsl}.

In this paper we show that biaxiality effects are closely related
to, but not exclusively confined to, the examples above. In fact,
within any spatially-varying director distribution, the director
gradient itself breaks uniaxial symmetry about the director. We
analyze in detail the structure of the elastic free energy density
and come up to the result that, given the director distribution, it
is possible to predict the onset of biaxiality, to determine the
direction of the secondary optic axis and to estimate the intensity
of biaxiality effects. We then apply our general considerations to
some specific examples, both within the bulk and close to an
external boundary. We remark that we are not dealing with
intrinsically biaxial nematic liquid crystals, that is systems in
which the ground state itself becomes biaxial. Such systems, first
observed by Yu and Saupe \cite{80yusa}, deserve a different
treatment \cite{95bich,03sovi}, since in them uniaxial symmetry is
broken already at a molecular level.

The paper is organized as follows. In Sections \ref{sec:ot} and
\ref{sec:fef} we quickly review the order-tensor theory and the free
energy density we aim at minimizing. In Section \ref{sec:bb} we
derive and describe our main result, predicting a possible onset of
biaxiality whenever the director is not uniform. In the following
Sections \ref{sec:ex} and \ref{sec:sb} we apply the preceding
results to some specific examples. In Section \ref{sec:disc} we
collect and discuss our main results, while some Appendices collect
the technical details of the proofs.

\section{Order tensor}\label{sec:ot}

The orientation of a single nematic molecule may be represented by a
unit vector $\nn\in\filS^2$, where $\filS^2$ is the unit sphere.
Microscopic disorder is taken into account by introducing a
probability measure $f_x:\filS^2\to\filR^+$, such that $f_x(\mm)$
describes the probability that a molecule placed in $x$ is oriented
along \mm. The probability measure $f_x$ is even, since opposite
orientations are physically equivalent.

Nematic optics is determined by the variance tensor
$\MM=\langle\,\mm\otimes\mm\,\rangle$, where the tensor product is
defined in (\ref{otimes}) and the brackets denote averaging with
respect to $f_x$. By definition, \MM\ is symmetric and semidefinite
positive. Since, in addition, the trace of \MM\ is equal to 1, we
define the traceless \emph{order tensor\/}
$\QQ=\MM-\frac{1}{3}\,\II$, where \II\ is the identity.

We label the nematic as \emph{isotropic\/} when all the eigenvalues
of \QQ\ coincide, which implies $\QQ_{\rm iso}=\mathbf{0}$. When at
least two eigenvalues are equal, the nematic is called
\emph{uniaxial\/}. Simple algebraic manipulations allow to write
\begin{equation}\label{quni}
\QQ_{\rm uni}=s\left(\nn\otimes\nn- \frac{1}{3}\,\II\right),
\end{equation}
The scalar parameter $s$ is the \emph{degree of orientation\/}
\cite{91er}, while the unit vector $\nn$ is the \emph{director}. The
eigenvalues of $\QQ_{\rm uni}$ are $\frac{2}{3}s$ (associated with
\nn) and $-\frac{1}{3}s$ (with multiplicity 2). The director is then
the eigenvector associated with the different eigenvalue.
Equivalently, \nn\ could be also identified as the eigenvector
associated with the eigenvalue whose sign is different from the
other two.

When the eigenvalues of the order tensor are all different, the
nematic is labeled as \emph{biaxial\/}. In this general case, we can
use the above remark, and still identify the director as the
eigenvector of \QQ\ whose eigenvalue has a different sign with
respect to the other two. This definition may induce an artificial
director discontinuity whenever the intermediate eigenvalue crosses
0. In turn, it yields an operative definition that works well when
the order tensor is possibly biaxial, but however close to being
uniaxial. Once we have introduced the director, we again define the
degree of orientation $s=\frac{3}{2}\lambda_{\rm n}$, where
$\lambda_{\rm n}$ is the eigenvalue associated with \nn. The other
two eigenvalues $\lambda_\pm$ can be finally written in terms of the
\emph{degree of biaxiality\/} $\beta$:
$\lambda_\pm=-\frac{1}{3}\,s\pm\beta$. As a result we obtain
\begin{equation}\label{qbia}
\QQ_{\rm bia}=s\left(\nn\otimes\nn- \frac{1}{3}\,\II\right)+\beta
\left(\epp\otimes\epp-\emm\otimes\emm\right).
\end{equation}
The sign of $\beta$ is unessential, since it only involves an
exchange between $\epp$ and $\emm$. The degree of biaxiality does
always satisfy $|\beta|\leq \frac{1}{3}|s|$. Indeed, when $|\beta|=
\frac{1}{3}|s|$ one of the eigenvalues vanishes, and greater
biaxiality values would in fact announce an abrupt change in the
director (and in the degree of orientation as well).

\section{Free energy functional}\label{sec:fef}

Equilibrium states of nematic liquid crystals are identified as
extremals of the free-energy functional whose density, in the
absence of external fields, comprises two terms
\begin{equation}\label{prfue}
\Psi(\QQ,\nabla\QQ) =\Psi_{\rm el}(\QQ,\nabla\QQ) + \Psi_{\rm
LdG}(\QQ)\;.
\end{equation}

Though all the calculations we report could be repeated in a more
general framework, we will adopt the 1-constant approximation for
the elastic contribution $\Psi_{\rm el}$
\begin{equation}\label{psiel}
\Psi_{\rm el}(\QQ,\nabla\QQ)=\frac{K}{2} \,|\nabla \QQ|^2 \,,
\end{equation}
where $K$ is an average elastic constant.

The Landau-de Gennes potential $\Psi_{\rm LdG}$ is a
tempera\-ture-dependent thermodynamic contribution that takes into
account the material tendency to spontaneously arrange in ordered or
disordered states:
\begin{equation}\label{psildg}
\Psi_{\rm LdG}(\QQ)=A\,\tr\QQ^2-B\,\tr\QQ^3+C\,\tr\QQ^4\;.
\end{equation}
The material parameter $C$ must be positive to keep the free-energy
functional bounded from below. The potential (\ref{psildg}) depends
only on the eigenvalues of \QQ, and penalizes biaxial states
\cite{93bica}. Insertion of (\ref{qbia}) into (\ref{psildg}) returns
\begin{align}
\Psi_{\rm
LdG}(s,\beta)&=\frac{2}{9}\left(Cs^4-Bs^3+3As^2\right)\nonumber \\
&+\frac{2}{9}\left(6Cs^2+9Bs+9A\right)\beta^2
+2C\beta^4\;.\label{psildg2}
\end{align}
Let $\alpha=3A/(Cs_0^2)$. The absolute minimum of $\Psi_{\rm LdG}$
is located at the uniaxial configuration $(s_0>0,\beta=0)$, provided
\begin{equation}
\alpha\in[-2,1]\quad{\rm and}\quad
B=\textstyle{\frac{2}{3}}\,Cs_0(\alpha+2)\;.
\end{equation}

When looking for minimizers of the free energy functional, we take
into account that Landau-de Gennes' contribution usually dominates
the elastic one. This approximation holds as long as we do not get
too close to a nematic defect. Indeed, experimental observations
confirm that neither $s$ nor $\beta$ depart easily from their
preferred values $(s_0,0)$.

We then envisage a two-step minimization. In the first step
$(s,\beta)$ are constrained to their optimal values. Minimization
proceeds exactly as in Frank's director theory and yields an optimal
distribution $\nn(\rr)$. In the second step, we fix the director
distribution and determine the perturbative corrections it induces
in the optimal values of the scalar order parameters. As a result,
we prove that non-uniform director configurations may induce a
nonzero degree of biaxiality, and a reduction in the degree of
orientation. As a by-product we determine how a non-zero director
gradient breaks the local axial symmetry induced by the director,
and which direction is chosen by most molecules (among those
orthogonal to \nn).

\section{Bulk biaxiality}\label{sec:bb}

We collect in the present section our main result. In order to ease
the reader we defer most of the technical proofs to the appendices
below. We assume that a specific director distribution $\nn(\rr)$
has been determined by minimizing Frank's free-energy functional,
constrained by suitable boundary conditions. The director
distribution may also take into account the effects of any possible
external field.

In Appendix \ref{app:dirgra} we prove the following decomposition
for the director gradient
\begin{align}
\grad \nn &= \lad \,\ed \tp \ed + \lat \,\et \tp \et + \big(\curl
\nn \vp \nn\big)\tp \nn \nonumber\\
&+\textstyle\frac{1}{2}\, (\nn \cdot \curl \nn) \,\WW(\nn)\,,
\label{gradn}
\end{align}
where $\WW(\nn)$ denotes the skew tensor associated with \nn\ (see
Appendix \ref{app:dirgra}). Furthermore, $\{\lad,\lat\}$,
$\{\ed,\et\}$ are respectively the eigenvalues and eigenvectors of
the symmetric part of $\GG=\grad \nn - (\grad \nn) \nn \tp \nn$, the
third eigenvector of $\sym\GG$ being \nn, with null eigenvalue. We
remark that
\begin{equation}
\dv\nn=\tr\grad\nn=\lad+\lat\;.
\end{equation}

Let $\Sb$ be the symmetric tensor $\Sb=(\grad \nn)\,(\grad \nn)^T$.
By virtue of (\ref{nunit}) the director \nn\ is an eigenvector of
\Sb\ (with null eigenvalue). In Appendix \ref{app:otgr} we prove
that the elastic free energy density may be given the following form
\begin{align}
\Psi_{\rm el}=K\Big[\textstyle\frac{1}{3}\,|\grad s|^2+ |\grad
\beta|^2+s^2\,|\grad \nn|^2 \nonumber\\
+\beta^2\Big(|\grad \nn|^2+4|(\grad\epp)^T\emm|^2\Big)\nonumber\\
-2s\beta\Big(\epp\cdot\Sb\epp-\emm\cdot\Sb\emm\Big)&\Big]\;.
\label{psi2}
\end{align}
Let us analyze in detail the different terms appearing in
(\ref{psi2}). The first two terms are trivial, since they simply
penalize spatial variations of the scalar order parameters. They
remind that, even in the presence of spatially-varying preferred
values $\big(s_{\rm opt}(\rr),\beta_{\rm opt}(\rr)\big)$, the
equilibrium distribution may not imitate the optimal values. The
third term is proportional to $s^2\,|\grad \nn|^2$. This term has
been already extensively studied \cite{91er,06bitu}. Its net effect
is a decrease in the degree of orientation in places where the
director gradient is most rapidly varying. In particular, it
strongly pushes the system towards the isotropic state $s=0$ when
the director gradient diverges. The second-last term is proportional
to $\beta^2$. Since it is positive-definite, it simply enhances the
character of $\beta=0$ as optimal biaxiality value. Thus, were not
for the final term we will next consider, biaxiality would never
arise naturally in a nematic liquid crystal.

The last term in (\ref{psi2}) is linear in $\beta$. It shifts the
optimal biaxiality value away from $\beta=0$. In order to minimize
the complete free energy density it is worth to maximize the
multiplying factor depending on \Sb. This condition determines the
directions $\{\epp,\emm\}$ in which the order tensor \QQ\ is pushed
to break uniaxial symmetry. Indeed, the term within brackets is
maximized when $\{\epp,\emm\}$ coincide with the two eigenvectors of
\Sb\ that are orthogonal to \nn. If we denote by $\mu_+,\mu_-$ the
correspondent eigenvalues, the linear term becomes simply
proportional to $(\mu_+-\mu_-)$. We thus arrive at the following
result. \emph{Consider the symmetric tensor $\Sb=(\grad \nn)\,(\grad
\nn)^T$. It always possesses a null eigenvalue (with eigenvector\/
\nn). Whenever its other two eigenvalues do not coincide, biaxiality
is naturally induced in the system, and the optimal eigendirections
of\/ \QQ\ coincide with those of\/ \Sb.\/}

If we take into account expression (\ref{gradn}) for $\grad\nn$, we
can give the eigenvalue difference $(\mu_+-\mu_-)$ the following
expression (see (\ref{appmupm}))
\begin{align}
(\mu_+-\mu_-)^2&=(c_2^2-c_3^2+\lambda_3^2-\lambda_2^2)^2\nonumber \\
&+ 4\left( c_2c_3+
\textstyle\frac{1}{2}c_n(\lambda_3-\lambda_2)\right)^2\,,\label{mupm}
\end{align}
where $\curl \nn=c_n\,\nn+c_2\,\ed+c_3\,\et$, and $\lad,\lat$ are as
in (\ref{gradn}). In the following sections we will apply the above
results to some practical situations, in order to better interpret
their implications.

\section{Splay, bend and twist biaxiality}\label{sec:ex}

\subsection{Pure splay}

We begin by considering the splay field $\nn(\rr)=\ee_r$, where
$\ee_r$ is the radial unit vector in cylindrical co-ordinates. If we
complete an orthonormal basis by introducing the tangential and
axial unit vectors $\ee_\theta,\ee_z$, standard calculations allow
to prove that the relevant fields are given by
\begin{align}
\grad\nn&=\frac{1}{r}\,\ee_\theta\tp\ee_\theta\quad
\Longrightarrow\quad\curl\nn=\bz\nonumber\\
\sym\GG&=\GG=\grad\nn=\frac{1}{r}\,\ee_\theta\tp\ee_\theta
\nonumber\\
\Sb &= (\grad \nn)^2 = \frac{1}{r^2}\,\ee_\theta\tp\ee_\theta
\nonumber\\
(\grad\epp)^T\emm&=(\grad\ee_\theta)^T\ee_z=\bz\;.
\end{align}
Thus, $\mu_+=r^{-2}$, $\mu_-=0$, and the elastic free energy density
is given by
\begin{equation}
\Psi_{\rm el}=K\left(\frac{1}{3}\,|\grad s|^2+ |\grad
\beta|^2+\frac{(s-\beta)^2}{r^2}\right)\;.\label{psplay}
\end{equation}
Biaxiality favours the tangential direction with respect to the
axial direction. The $r^{-2}$ factor implies that biaxiality (and
the degree of orientation decrease as well) is expected to show
close to the symmetry axis. Figures 3 and 5 of \cite{97bivi} exactly
confirm this result.

\subsection{Pure bend}

We again consider the same cylindrical coordinate frame above, and
analyze the bend field $\nn(\rr)=\ee_\theta$. We obtain
\begin{align}
\grad\nn&=-\frac{1}{r}\,\ee_r\tp\ee_\theta\quad
\Longrightarrow\quad\curl\nn=\frac{1}{r}\,\ee_z\nonumber\\
\sym\GG&=\GG=\bz\nonumber\\
\Sb &= \frac{1}{r^2}\,\ee_r\tp\ee_r\nonumber\\
(\grad\epp)^T\emm&=(\grad\ee_r)^T\ee_z=\bz\;.
\end{align}
Again, $\mu_+=r^{-2}$, $\mu_-=0$, and the elastic free energy
density can given exactly the same expression (\ref{psplay}).
Biaxiality now favours the radial direction, and again concentrates
close to the (disclination) symmetry axis.

\subsection{Pure twist}

We now introduce a Cartesian frame $\{\ee_x,\ee_y,\ee_z\}$ and
consider the twist field $\nn(\rr)=\cos kz\,\ee_x+ \sin kz\,\ee_y$.
If we introduce the unit vector $\nn_\perp(\rr)=-\sin kz\,\ee_x+
\cos kz\,\ee_y$, we obtain
\begin{align}
\grad\nn&=k\,\nn_\perp\otimes\ee_z\quad
\Longrightarrow\quad\curl\nn=-k\,\nn\nonumber\\
\GG&=\grad\nn\quad \Longrightarrow\quad
\sym\GG=k\,\sym\nn_\perp\otimes
\ee_z\nonumber\\
\Sb &= k^2\,\nn_\perp\otimes\nn_\perp\nonumber\\
(\grad\epp)^T\emm&=(\grad\nn_\perp)^T\ee_z=\bz\;.
\end{align}
We now have $\mu_+=k^2$, $\mu_-=0$. Biaxiality favours $\nn_\perp$,
that is, the $(x,y)$ plane, with respect to the transverse direction
$\ee_z$. The elastic free energy density does again coincide with
(\ref{psplay}), with only a $k^2$ replacing the $r^{-2}$ factor.
However, this coincidence must not induce to guess that $\Psi_{\rm
el}$ does always depend on $s$ and $\beta$ only through the
combination $(s-\beta)$, as we will evidence below.

\subsection{Third-dimension escape}

All the model cases analyzed above share a peculiar property.
Indeed, the fields $\nn(\rr)$ considered are all planar, in that
they are all orthogonal to a fixed direction ($z$-axis). Therefore
it is of no surprise that we always find some degree of biaxiality
which penalizes the avoided direction. We now consider a less
trivial example: the escape in the third-dimension. This field was
first determined by Cladis and Kl\'eman \cite{72clkl} as an
everywhere continuous director field able to fulfill homeotropic
boundary conditions on a cylinder. Let
$\nn(\rr)=\cos\phi(r)\,\ee_r+\sin\phi(r)\,\ee_z$ be the director
field, and let
$\nn_\perp(\rr)=-\sin\phi(r)\,\ee_r+\cos\phi(r)\,\ee_z$. We obtain
\begin{align}
&\grad\nn=\frac{\cos\phi}{r}\,\ee_\theta\tp\ee_\theta+
\phi'\,\nn_\perp\tp\ee_r \nonumber\\
&\curl\nn=- \phi'\cos\phi\,\ee_\theta\nonumber \\
&\sym\GG=\GG=\frac{\cos\phi}{r}\,\ee_\theta\tp\ee_\theta -
\phi'\sin\phi\,\nn_\perp\tp\nn_\perp\nonumber\\
&\Sb =\frac{\cos^2\phi}{r^2}\,\ee_\theta\tp\ee_\theta + \phi^{\prime
2}\,\nn_\perp\tp\nn_\perp\;.\label{3rd}
\end{align}
Expression (\ref{3rd}) for \Sb\ shows that, within the order tensor
\QQ, either $\nn_\perp$ or $\ee_\theta$ may be preferred, depending
on whether $\phi^{\prime 2}$ is greater or smaller than
$\cos^2\phi/r^2$. This result turns out to be particularly
challenging, if we consider that in Cladis-Kl\'eman's escape in the
third dimension the tilt angle $\phi$ is given by
\begin{equation}
\phi(r)=\frac{\pi}{2}-2\arctan\frac{r}{R}\;.\label{clkl}
\end{equation}
A simple calculation allows to show that (\ref{clkl}) implies
$\phi^{\prime 2}=\cos^2\phi/r^2$. Thus, the third-dimension escape
turns out to be one of the few spatially-varying director fields
which do not induce any biaxiality. The elastic free-energy density
in Cladis-Kleman's third-dimension escape is given by
\begin{align}
\Psi_{\rm el}=K\left(\frac{1}{3}\,|\grad s|^2+ |\grad
\beta|^2\right.&+\frac{8R^2\,s^2}{(r^2+R^2)^2}\nonumber\\
&\left.+\frac{4(R^4+r^4)\,\beta^2}{r^2(r^2+R^2)^2}\right)\;.
\label{psi3rd}
\end{align}

\section{Surface biaxiality}\label{sec:sb}

In this section we estimate the degree of biaxiality induced by an
external surface on which strong anchoring is enforced. We consider
separately the cases of homeotropic and planar anchoring.
Differential calculus formulae that turn out to be useful for both
cases are collected in Appendix \ref{app:surf}.

\subsection{Homeotropic anchoring}

\begin{figure}
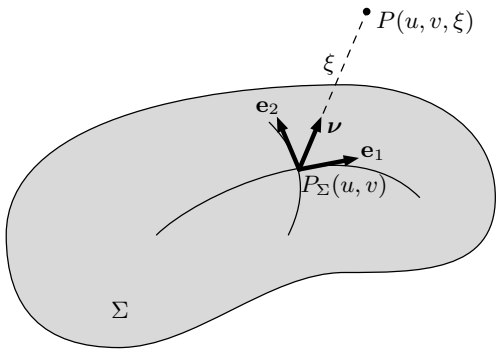

\begin{texdraw}
\drawdim cm \setunitscale 1.0
\linewd 0.02
\move (0 0) \clvec (1 0)(2 1)(3 1)
\clvec (4 1)(5 1)(5 2) \clvec (5 3)(4 3.5)(3 3.5) \clvec (1
3.5)(-1.5 3)(-1.5 1.5) \clvec (-1.5 0.5)(-1 0)(0 0) \lfill f:0.85
\move (0.5 1.5) \clvec (1 2)(3 3)(4 2)
\move (2 3) \clvec (2.5 2.5)(2.5 2)(2.25 1.5)
\arrowheadtype t:F \arrowheadsize l:.2 w:.15
\linewd 0.06
\move (2.4 2.37) \ravec (0.8 0.15)
\move (2.4 2.37) \ravec (-0.3 0.7)
\move (2.4 2.37) \ravec (0.3 0.7)
\linewd 0.02 \lpatt(0.1 0.1)
\move (2.4 2.37) \rlvec (0.9 2.1) \fcir f:0 r:0.05
\lpatt ()
\textref h:L v:B \htext (3.2 2.5) {$\ee_1$}
\textref h:L v:B \htext (1.8 3.1) {$\ee_2$}
\textref h:L v:T \htext (2.75 3) {\hbox{\bnu}}
\textref h:L v:T \htext (2.7 3.9) {$\xi$}
\textref h:L v:T \htext (3.4 4.5) {$P(u,v,\xi)$}
\textref h:L v:T \htext (2.4 2.3) {$P_\Sigma(u,v)$}
\textref h:C v:C \htext (0 0.5) {$\Sigma$}
\end{texdraw}
\caption{Geometric setting for the surface parametrization
introduced in the text.}\label{figuno}
\end{figure}

We first assume that the surface director is parallel to the unit
normal $\bnu$ to a given (smooth) surface $\Sigma$. We also assume
that the director keeps its normal direction, at least in a thin
surface slab. To be more precise, we parameterize bulk points
through a coordinate set $(u,v,\xi)$ such that
\begin{equation}
P(u,v,\xi)=P_\Sigma(u,v)+\xi\,\bnu(u,v)\,,
\end{equation}
where $P_\Sigma$ is the projection of $P$ onto $\Sigma$, $\xi$ is
the distance of $P$ from the fixed surface, and $\bnu$ is the unit
normal at $P_\Sigma$ (see Figure \ref{figuno}). If $\Sigma$ if
smooth, the coordinate set is well-defined in a finite neighborhood
of $\Sigma$. We assume that
$\nn\big(P(u,v,\xi)\big)=\nn\big(P_\Sigma(u,v)\big)=\bnu(u,v)$.
Then, $\grad\nn$ turns out to be closely related to the curvature
tensor. It is symmetric and can be written as (see (\ref{cc2}))
\begin{equation}
\grad \nn = -\frac{\kuno}{1-\kuno \xi} \, \ee_1 \tp \ee_1 -
\frac{\kdue}{1-\kdue \xi}\, \ee_2 \tp \ee_2\;,\label{gns}
\end{equation}
where $\{\kappa_1,\kappa_2\}$ and $\{\ee_1,\ee_2\}$ denote
respectively the principal curvatures and principal directions at
$P_\Sigma$. From them we obtain $\curl\nn=\bz$, $\GG=\grad\nn$, and
\begin{equation}
\Sb=\frac{\kappa_1^2}{(1-\kuno \xi)^2} \, \ee_1 \tp \ee_1+
\frac{\kappa_2^2}{(1-\kdue \xi)^2}\, \ee_2 \tp \ee_2\;.\label{sbs}
\end{equation}
Equation (\ref{sbs}) shows that biaxiality arises naturally close to
an external surface where homeotropic anchoring is enforced. This
effect is triggered by the difference between the principal
curvatures. More precisely, the tangent direction preferred by the
order tensor is the one along which the surface curves more rapidly.
Close to a symmetric saddle, where $\kappa_1=-\kappa_2$, the
denominator of (\ref{sbs}) induces biaxiality along the direction
which is convex towards the side occupied by the liquid crystal.

\subsection{Planar anchoring}

When planar anchoring is enforced on a curved surface, it is natural
to assume that the chosen direction coincides with one of the
principal directions along $\Sigma$. We then keep the same notations
as above, and assume
$\nn\big(P(u,v,\xi)\big)=\nn\big(P_\Sigma(u,v)\big)=\ee_1(u,v)$.
When this is the case, by (\ref{cc2}) we obtain
\begin{align}
\grad \nn&= \frac{\kuno}{1-\kuno \xi} \, \bnu \tp
\ee_1\quad\Longrightarrow\quad\curl \nn=-\frac{\kuno\,\ee_2}
{1-\kuno \xi}\nonumber\\
\GG&=\bz\quad{\rm and}\quad \Sb=\frac{\kappa_1^2}{(1-\kuno \xi)^2}
\; \bnu \tp \bnu\;.\label{planar}
\end{align}
Thus, in the presence of planar anchoring, biaxiality arises
whenever the curvature along the prescribed direction is different
from zero. When this is the case, the biaxiality direction coincides
with the unit normal.

\section{Discussion}\label{sec:disc}

We have shown that any spatially-varying director distribution may
induce the onset in biaxial domains even in nematic liquid crystals
whose ground state is strictly uniaxial. In particular, in Section
\ref{sec:bb} we have stressed the crucial role played by
$\Sb=(\grad\nn)(\grad\nn)^T$. The tensor $\Sb$, which is symmetric
and positive-semidefinite by construction, possesses always a null
eigenvalue, with eigenvector \nn. Equation (\ref{psi2}) shows that
biaxiality arises naturally whenever the other two eigenvalues of
\Sb\ are different. Then, equation (\ref{mupm}) shows that such
eventuality is closely related to the vector $\curl\nn$ and the
eigenvalues entering in the decomposition (\ref{gradn}) of the
director gradient.

In Section \ref{sec:ex} we have applied the considerations above to
some model cases. As it could be easily predicted the pure splay,
bend, and twist fields, being all planar, exhibit some degree of
biaxiality which privileges the director plane over the orthogonal
director. A less trivial result is that there are spatially-varying
director configurations that do not induce biaxiality at all.
Cladis-Kl\'eman's escape in the third dimension yields a unexpected
example of this phenomenon. Section \ref{sec:sb} analyzes the onset
of surface biaxiality both in the case of homeotropic and planar
alignment. In the former case, biaxiality is ruled by the difference
between the principal curvatures along the surface. In the latter,
only one curvature counts, and more precisely the one along the
prescribed direction in the tangent plane.

To conclude our analysis we want to give a numerical estimate of the
magnitude of the biaxiality phenomena we are predicting. In all
nontrivial cases, the free-energy density will contain a
$O(\beta)$-term, which triggers the biaxiality onset. To obtain a
rough estimate, we can neglect the $O(\beta^4)$-term in $\Psi_{\rm
LdG}$, and the $O(\beta^2)$-term in $\Psi_{\rm el}$, both with
respect to the dominant $O(\beta^2)$-term, appearing in $\Psi_{\rm
LdG}$. When this is the case, the (local) preferred value of $\beta$
may be obtained by minimizing the function
\begin{align}
g(\beta)&=\frac{2}{9}\left(6Cs^2+9Bs+9A\right)\beta^2
-2Ks\beta\big(\mu_+-\mu_-\big)\nonumber\\
&\approx2(2+\alpha)Cs_0^2\beta^2 -2Ks_0\beta\big(\mu_+-\mu_-\big)
\nonumber\\
&=\frac{2Ks_0}{\xi_{\rm n}^2}\Big(\beta^2 -\xi_{\rm
n}^2\big(\mu_+-\mu_-\big)\beta\Big)\,,
\end{align}
where we have replaced $s\approx s_0$ and introduced the
\emph{nematic coherence length\/}
\begin{equation}
\xi_{\rm n}^2=\frac{K}{Cs_0(2+\alpha)}\;.
\end{equation}
The (local) optimal value of the degree of biaxiality is then
\begin{equation}
\beta_{\rm opt}\approx\textstyle\frac{1}{2}\,\xi_{\rm
n}^2\,\big(\mu_+-\mu_-\big)\;.\label{beopt}
\end{equation}
Though $\beta_{\rm opt}$ may vary from point to point, we have to
keep in mind that in general the equilibrium configuration will not
coincide with $\beta_{\rm opt}$ because of the
$|\grad\beta|^2$-term, and the boundary conditions. To make an
explicit example, let us consider a nematic cylindric capillary of
radius $R$, with homeotropic conditions enforced at the surface.
Then, the difference between the eigenvalues of $\Sb$ at the surface
is $R^{-2}$ and the surface biaxiality is of the order of $(\xi_{\rm
n}/R)^2$. Since the nematic coherence length hardly exceeds the
tenths of a $\mu$m, we obtain $\beta_{\rm opt}\lesssim 10^{-2}$ for
a $\mu$m-capillary. The scenario changes completely close to a
nematic defect, where at least one of the eigenvalues of \Sb\
diverges. Both the pure-splay and the pure-bend examples above yield
$(\mu_+-\mu_-)=r^{-2}$, which implies $\beta_{\rm opt}\approx
(\xi_{\rm n}/r)^2$. The biaxiality \emph{cloud\/} cannot be
neglected if we come too close to the defect.

\appendix

\section{Director gradient}\label{app:dirgra}

In order to characterize the tensor $\grad\nn$ we begin by noticing
that
\begin{equation}
(\grad \nn)^T \nn
=\frac{1}{2}\grad\big(\nn\cdot\nn\big)=\bz\;,\label{nunit}
\end{equation}
since \nn\ is a unit vector. Thus,
\begin{align}
\nonumber (\grad \nn) \nn &= \big(\sym \grad \nn + \skw \grad
\nn\big)\nn \\
&= \frac{1}{2}\, (\grad \nn) \nn + \frac{1}{2} \,\curl \nn \vp
\nn\,,\label{aa2}
\end{align}
and  $(\grad \nn) \nn = \curl \nn \vp \nn$. Let $\GG = \grad \nn -
(\grad \nn) \nn \tp \nn$. For any vector \vv,
\begin{align}
(\skw \GG) \vv &= \left(\skw( \grad \nn) -
\textstyle\frac{1}{2}[(\grad \nn) \nn \tp
\nn - \nn \tp (\grad \nn) \nn]\right)\vv \nonumber \\
&= \textstyle\frac{1}{2} \Big(\curl \nn -
\nn\vp (\curl \nn \vp \nn)\Big)\vp\vv \label{app3}\\
&= \textstyle\frac{1}{2} (\nn \cdot \curl \nn) \nn \vp \vv =
\frac{1}{2} (\nn \cdot \curl \nn) \WW(\nn) \vv\,,\nonumber
\end{align}
where $\WW(\nn)$ denotes the skew tensor associated with \nn, that
is the tensor such that $\WW(\nn)\vv=\nn\vp\vv$ for any \vv.
Thus,
\begin{align}
\grad \nn &= \sym \GG + \textstyle\frac{1}{2} (\nn \cdot \curl \nn)
\,\WW(\nn) + \big(\curl \nn \vp \nn\big)\tp \nn \nonumber \\
&=\lad \ed \tp \ed + \lat \et \tp \et+ \textstyle\frac{1}{2} (\nn
\cdot \curl \nn) \WW(\nn)\nonumber\\
&+ \big(\curl \nn \vp \nn\big)\tp \nn\,,\label{agrad}
\end{align}
where $\{\lad,\lat\}$ and $\{\ed,\et\}$ are respectively the
eigenvalues and eigenvectors of $\sym\GG$. The eigenvectors
$\{\ed,\et\}$ are orthogonal to \nn, since (\ref{app3}) implies
\begin{align}
(\sym\GG)\nn&=\GG\nn-(\skw \GG)\nn=\big(\grad \nn - (\grad \nn) \nn
\tp \nn\big)\nn\nonumber\\
&- \frac{1}{2} (\nn \cdot \curl \nn) \WW(\nn) \nn=\bz\;.
\end{align}
Let us consider the symmetric tensor $\Sb=(\grad \nn)(\grad \nn)^T$.
In view of the crucial role it plays in inducing biaxiality we now
analyze it in more detail
\begin{align}
\Sb&=\big[\lad \ed \tp \ed + \lat \et \tp \et+ \textstyle\frac{1}{2}
(\nn\cdot \curl \nn) \WW(\nn)\nonumber\\
&+ \big(\curl \nn \vp \nn\big)\tp \nn\big]\big[\lad \ed \tp \ed +
\lat \et \tp \et\nonumber\\
&- \textstyle\frac{1}{2} (\nn\cdot \curl \nn) \WW(\nn)+ \nn\tp
\big(\curl \nn \vp \nn\big)\big]\nonumber\\
&=\lambda_2^2 \ed \tp \ed +
(\lad-\lat) (\nn\cdot \curl \nn) \sym(\ed\tp\et)\nonumber\\
&\nonumber + \lambda_3^2 \et \tp \et+\textstyle\frac{1}{4} (\nn\cdot
\curl \nn)^2 (\II-\nn\tp\nn) \\
&+ \big(\curl \nn \vp \nn\big)\tp \big(\curl \nn \vp \nn\big)\;.
\label{appSb}
\end{align}
Let $\{0,\mu_+,\mu_-\}$ be the eigenvalues of $\Sb$. The onset of
biaxiality depends whether they latter two are equal or not. Let
$\curl \nn=c_n\,\nn+c_2\,\ed+c_3\,\et$. From (\ref{appSb}) we obtain
\begin{align}
(\mu_+-\mu_-)^2&=(c_2^2-c_3^2+\lambda_3^2-\lambda_2^2)^2\nonumber \\
&+ 4\left( c_2c_3+
\textstyle\frac{1}{2}c_n(\lambda_3-\lambda_2)\right)^2\,.\label{appmupm}
\end{align}

We finally remind that, given two vectors \uu,\vv, the tensor
product $(\uu\otimes\vv)$ is defined as the second order tensor such
that
\begin{equation}
(\uu\otimes\vv)\ab=(\vv\cdot\ab)\,\uu\quad\text{for any vector
}\ab\,. \label{otimes}
\end{equation}

\section{Order tensor gradient}\label{app:otgr}

Let us differentiate equation (\ref{qbia}). We obtain
\begin{align}
\grad\QQ&=\left(\nn\otimes\nn-\textstyle{\frac{1}{3}}\II\right)
\otimes\grad s + s \left( \grad \nn\odot \nn +\nn\otimes\grad
\nn\right)\nonumber\\
&+\left(\epp\otimes\epp-\emm\otimes\emm\right)\otimes\grad\beta
+\beta \big( \grad \epp\odot \epp\nonumber\\
&+\epp\otimes\grad \epp -\grad \emm\odot \emm-\emm\otimes\grad
\emm\big)\,,
\end{align}
where, given the second order tensor \LL\ and the vector \uu,
$(\LL\odot\uu)$ is defined as the third order tensor such that
\begin{equation}
(\LL\odot\uu)\ab=\LL\ab\otimes\uu\quad\text{for any vector }\ab\,.
\end{equation}
When computing the square norm of $\grad\QQ$, we can make extensive
use of the property (\ref{nunit}) and also take into account that
\begin{equation}
\uu\cdot\vv=0\quad\Longrightarrow\quad (\grad \uu)^T\vv=- (\grad
\vv)^T\uu\;.
\end{equation}
As a consequence, we obtain
\begin{align}
|\grad\QQ|^2&=\textstyle\frac{2}{3}\,|\grad s|^2+2 \,|\grad
\beta|^2+2s^2\,|\grad \nn|^2 \nonumber\\
&+2\beta^2\Big(|\grad \epp|^2+|\grad \emm|^2+2|(\grad\epp)^T\emm|^2\Big)\nonumber\\
&-4s\beta\Big(|(\grad\nn)^T\epp|^2-|(\grad\nn)^T\emm|^2\Big)\;.
\label{med0}
\end{align}
We can further simplify expression (\ref{med0}) if we consider that
\begin{align}
|\grad \epp|^2+|\grad \emm|^2&=\big|(\grad \epp)^T\big|^2
+\big|(\grad \emm)^T\big|^2\nonumber\\
&=\big|(\grad \epp)^T\nn\big|^2 +\big|(\grad \epp)^T\emm \big|^2
\nonumber\\
&+\big|(\grad \emm)^T\nn\big|^2 +\big|(\grad \emm)^T\epp \big|^2
\nonumber\\
&=\big|\grad \nn\big|^2 +2\big|(\grad \epp)^T\emm \big|^2
\label{med1}
\end{align}
and that
\begin{equation}
|(\grad\nn)^T\uu|^2=(\grad\nn)^T\uu\cdot(\grad\nn)^T\uu=
\uu\cdot\Sb\uu\,,\label{med2}
\end{equation}
provided we define $\Sb=(\grad\nn)(\grad\nn)^T$. By using
(\ref{med1})-(\ref{med2}) it is immediate to give (\ref{med0}) the
expression quoted in (\ref{psi2}).

\section{Curvature tensor}\label{app:surf}

Let $\Sigma$ be the smooth surface, which bounds the system we are
interested in. Let $\bnu$ be the unit normal, everywhere pointing in
the direction of the bulk. We parameterize points in the bulk
through a coordinate set $(u,v,\xi)$ such that
\begin{equation}
P(u,v,\xi)=P_\Sigma(u,v)+\xi\,\bnu(u,v)\,,
\end{equation}
where $P_\Sigma$ is the projection of $P$ onto $\Sigma$, and $\xi$
is the distance of $P$ from the same surface. Such coordinate set is
well-defined in a finite neighborhood of $\Sigma$.

Let us consider the vector field everywhere defined as
$\btau\big(P(u,v,\xi)\big)=\bnu\big(P_\Sigma(u,v)\big)$. The
second-order tensor $\grad\btau$ is symmetric. It generalizes the
\emph{curvature tensor\/} $\grad_\Sigma\bnu$, which is defined only
on the tangent bundle of $\Sigma$. The eigenvectors of $\grad\btau$
coincide with those of the curvature tensor, and are thus the unit
normal (with null eigenvalue) and the (tangent) \emph{principal
directions\/} on $\Sigma$. If we introduce $\{\kappa_1,\kappa_2\}$,
the \emph{principal curvatures\/} on $\Sigma$, and their
corresponding eigenvectors $\{\ee_1,\ee_2\}$, we have
\begin{align}
\grad \btau &= -\frac{\kuno}{1-\kuno \xi} \, \ee_1 \tp \ee_1 -
\frac{\kdue}{1-\kdue \xi}\, \ee_2 \tp \ee_2\label{cc2}\\
\grad \ee_1&= \frac{\kuno}{1-\kuno \xi} \, \bnu \tp \ee_1\quad{\rm
and} \quad \grad \ee_2 = \frac{\kdue}{1-\kdue \xi} \, \bnu \tp
\ee_2\;.\nonumber
\end{align}

\end{document}